\documentclass[12pt,onecolumn,amsfonts]{revtex4-1}
\usepackage{amsfonts}
\usepackage{amsmath}
\usepackage{amssymb}
\usepackage{graphics,graphicx}
\usepackage{epsf}
\usepackage{hyperref}

\newcommand{\be}{\begin{equation}}
\newcommand{\ee}{\end{equation}}
\newcommand{\bem}{\begin{multline}}
\begin{document}

%\preprint{}

\title{Structure, dynamics and bifurcations of discrete solitons in trapped ion crystals}

\author{H. Landa}
\email{haggaila@gmail.com}
\author{B. Reznik}
\affiliation{School of Physics and Astronomy, Raymond and Beverly Sackler Faculty of Exact Sciences, Tel-Aviv University, Tel-Aviv 69978, Israel}

\author{J. Brox}
\author{M. Mielenz}
%\author{S. Kahra}
%\author{G. Leschhorn}
%\author{M. Albert}
\author{T. Schaetz}
\affiliation{Albert-Ludwigs-Universit\"at Freiburg, Physikalisches Institut, Hermann-Herder-Strasse 3, 79104 Freiburg, Germany}
\affiliation{Max-Planck-Institut f\"ur Quantenoptik, Hans-Kopfermann-Strasse 1, D-85748 Garching, Germany}

\begin{abstract}
We study discrete solitons (kinks) accessible in state-of-the-art trapped ion experiments, considering zigzag crystals and quasi-3D configurations, both theoretically and experimentally. We first extend the theoretical understanding of different phenomena predicted and recently experimentally observed in the structure and dynamics of these topological excitations. Employing tools from topological degree theory, we analyze bifurcations of crystal configurations in dependence on the trapping parameters, and investigate the formation of kink configurations and the transformations of kinks between different structures. This allows us to accurately define and calculate the effective potential experienced by solitons within the Wigner crystal, and study how this (so-called Peierls-Nabarro) potential gets modified to a nonperiodic globally trapping potential in certain parameter regimes. The kinks' rest mass (energy) and spectrum of modes are computed and the dynamics of linear and nonlinear kink oscillations are analyzed. We also present novel, experimentally observed, configurations of kinks incorporating a large-mass defect realized by an embedded molecular ion, and of pairs of interacting kinks stable for long times, offering the perspective for exploring and exploiting complex collective nonlinear excitations, controllable on the quantum level.
\end{abstract}


\maketitle
\tableofcontents
\section{Introduction\label{Intro}}

The suggestion to investigate discrete solitons (kinks) in trapped ion Wigner crystals has been put forward in \cite{KinkCoherence}, where structural and spectral properties of different types of topological defects were studied, and an analysis of the quantization of modes of discrete solitons was presented. Such discrete kinks were first realized in \cite{Schneider2012}, and recently they were further studied both experimentally and theoretically in \cite{kink_trapping}, where novel types of kinks with intriguing characteristics were discovered. The Kibble-Zurek mechanism \cite{Kibble1976,Kibble1980,Zurek1985,Zurek1996} which allows to predict the density of structural defects formed during a second-order phase-transition, was analyzed for the finite and inhomogenous ion trap setting \cite{StructuralDefects,SpontaneousNucleation}, and was recently studied in experiments observing the formation of kinks with trapped ions \cite{TanjaKZ,MainzKZ,HaljanKinks}.

A basic property of solitons is their particle-like mobility. Understanding transport properties of discrete solitons is of importance for studying nonequilibrium processes in many nonlinear models widespread in physics \cite{BraunKivsharBook}, where discrete solitons (and more generally, discrete breathers \cite{Flach2008Review}) form basic excitations of the system. Even in other fields such as chemistry and biology, discrete solitons have been proposed for describing dynamical processes in, e.g., DNA and biological macromolecules, or in chains of ions or molecules interacting through hydrogen-bonds, where transfer of protons is described by propagating solitons \cite{dauxois2006physics}.

One of the most important concepts with respect to propagating discrete solitons is the effective potential landscape that the kinks moving in a lattice are subject to, known as the Peierls-Nabarro (PN) potential \cite{BraunKivsharBook}. The PN potential is typically periodic and can often lead to pinning of solitons in lattice sites, in particular because travelling solitons can radiate phonons and slow down. However, recently there is a rising interest in the study of novel phenomena associated with the PN potential in different models and schemes. Enhanced mobility of localized optical excitations in a coupled array of waveguides was discovered in \cite{PhysRevE.67.056606}, in connection with bifurcations of kink configurations. Exact vanishing of the effective PN potential for specific parameter values and velocities of the propagating solitons was found in some models corresponding to nonlinear optical media, in 1D and 2D \cite{PhysRevLett.93.033901,PhysRevLett.97.124101,PhysRevE.73.046602,PhysRevE.76.036603,PhysRevLett.99.214103,PhysRevE.83.036601}. With discrete solitons of cold condensed atoms in optical lattices, mobility within the PN potential has also been studied \cite{PhysRevA.69.053604}, and possibilities of controlled `routing' of matter-wave solitons have been suggested \cite{malomed2006soliton,PhysRevLett.105.090401}.

Cold trapped ions offer a high degree of experimental control \cite{Schneider2012} allowing to study the nonlinear physics of discrete solitons, possbily even in the quantum regime \cite{KinkCoherence}, which is usually experimentally inaccessible. Discrete solitons have also been recently proposed for high-fidelity entanglement generation in ion traps \cite{Marcovitch,soliton_gate}, at the presence of noise and with prospects for studying open quantum system dynamics. Such applications and further fundamental studies of solitons in ion traps could benefit substantially from accurate control over the solitonic motional degrees of freedom, and in particular, from a deeper understanding of soliton mobility dynamics in the lattice.

In this paper we study the structure and dynamics of discrete solitons in inhomogenous, trapped Wigner crystals. In particular we focus on the formation and properties of the PN potential as a function of trap parameters for various kink types. We analyze the bifurcations in the space of crystal configurations, which lead to formation of the PN potential. We investigate further the unique phenomenon of the PN potential becoming a global trapping potential for certain kinks, with strong anharmonic modification of the kink motion. We study, both experimentally and theoretically, single-kink configurations, two-kink interactions, and a novel type of kink, modified by a large-mass defect.

\section{Bifurcations in Trapped Ion Crystals\label{Bifurcations}}

In this section we study theoretically  different kink solutions with 31 ions, restricted to a planar structure, where a single parameter controls the model, namely the ratio of the radial and axial confinement strength. We investigate the structure of the solution space of crystal configurations, and study how kinks of different types are related with other solutions. We analyze bifurcations which lead to kink solutions as the model parameter is varied, and explain their structure from considerations of symmetry and topology. All of these types of kinks were observed experimentally.

Crystals with $N$ ions of a single species trapped in a quadrupole trap \cite{WinelandReview}, can be described within the pseudopotential approximation by using the nondimensional potential energy in the form 
\be {V} = \sum _{i}^{N}\frac{1}{2} \left(x_{i} ^{2} +\omega_{y}^2 y_{i} ^{2} +\omega_{z}^2 z_{i} ^{2} \right) + \sum _{i\ne j}\frac{1}{2} \left\| \vec{R}_{i} -\vec{R}_{j} \right\| ^{-1}, \label{Eq:VNondimful}\ee
where $\vec{R}_{i} = \left\{x_{i},y_{i},z_{i}\right\}$ is the vector coordinate of ion $i$, time and distances are measured in units of $1/{\omega_x}$ (with $\omega_x$ the axial trapping frequency) and $d=\left(e^{2} /m{\omega_x}^{2} \right)^{1/3}$, respectively (with $m$ and $e$ the ion's mass and charge), and the energy is in units of $m\omega_x^2d^2$.

For a small number of ions and if the axial confinement is much smaller than the radial, sufficiently cooled ions in a quadrupole trap form a crystalline chain along the trap symmetry axis. For a given number of ions and with lower radial confinement, the linear chain configuration becomes unstable, and the global minimum configuration takes the shape of a zigzag \cite{birkl1992zigzag,PhysRevLett.70.818,ZigzagTransitionExp,Peeters1D,MorigiFishman2008structural,ZigzagTransitionThermal,HaljanHopping,zigzagexperiment,molecular_conveyor_belt} which extends from the center of the trap, with possibly a part of the ions on the sides still in an axial chain. For parameters close enough to the transition from a linear chain, the zigzag is a planar structure. For radial confinement which is further decreased, a transition from the zigzag to a helix occurs for (nearly) degenerate radial frequencies \cite{birkl1992zigzag,PhysRevLett.70.818,Hasse1990419,Coulomb_double_helical,nigmatullin2011formation}, while planar and 3D structures can also form for suitable trapping parameters (see e.g. \cite{Mitchell13111998,Surface_Trap_2D_Crystals,Bermudez_spin_ladder,PhysRevLett.91.095002,PhysRevLett.96.103001}).

\begin{figure}[ht]
\center {\includegraphics[width=6in]{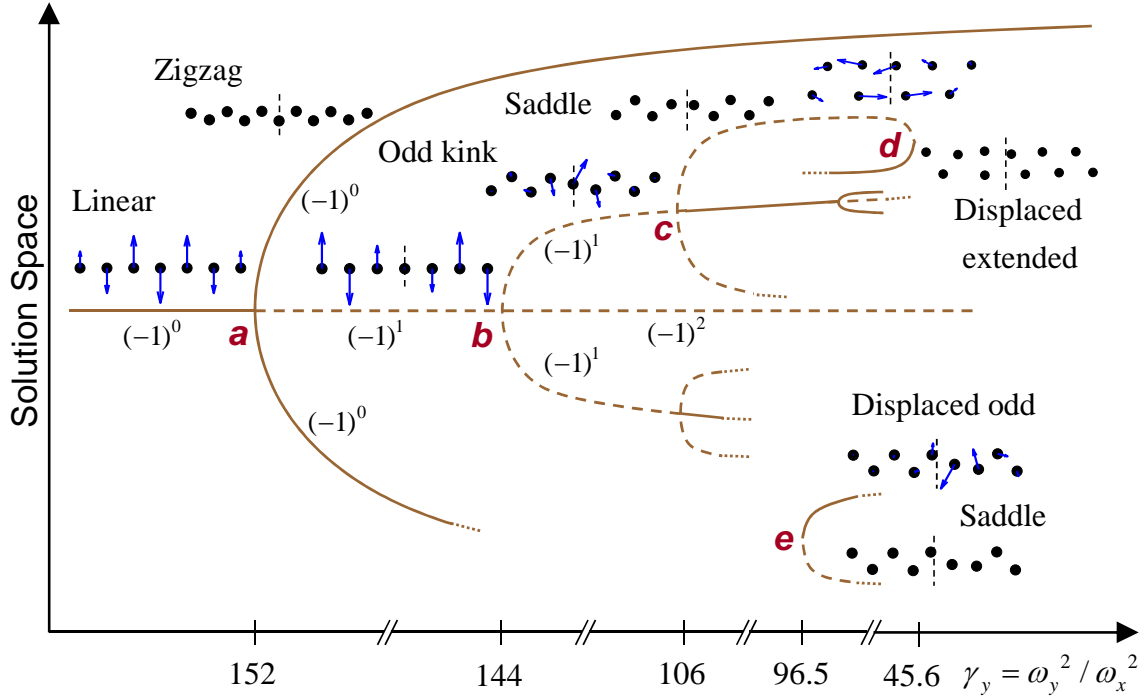}
\caption
{Bifurcations and stability of one-kink configurations in a planar crystal of 31 ions. The solution space of possible ion configurations is depicted schematically vs.~the single parameter which determines the structure of planar crystals, $\gamma_y\equiv \omega_y^2/\omega_x^2$. At each parameter value there are many possible crystal configurations, and a few of the these are indicated in the figure, which shows how they depend continuously on the model parameter. Solid lines indicate stable configurations (local minima of the potential) and dashed lines are unstable solutions, with at least one negative eigenvalue of the Hessian matrix $K$. The number of negative eigenvalues $n$ is indicated (next to selected solutions) in the form $\left(-1\right)^n$, which gives the local index of the Hessian at each configuration, $i_0\equiv {\rm sign}\det K$. The sum of local indices of all solutions is conserved on the left and right of a bifurcation. The spatial configurations are depicted by showing a few ions at the crystal centre, together with the line of axial symmetry of the crystal, and the arrows indicate the direction and amplitude of ion motion in the normal mode which crosses zero at the bifurcation and generates the bifurcating solutions. See text for details of the bifurcations \textit{a}-\textit{e}.}
\label{Fig:Bifurcations}}
\end{figure}

In this section we focus on planar crystals in $xy$ plane, i.e. such that $\omega_z \gg \omega_y$, omitting the dependence on the $z$ coordinate from the discussion. Within this description the structure of the planar crystals is therefore determined by the single parameter $\gamma_y \equiv \omega_y^2/\omega_x^2$. Let us define a crystal configuration as any critical point of the potential of eq.~\eqref{Eq:VNondimful}, i.e. a solution of the equation $\nabla V =0$. In other words, the static configurations are zeros of the vector field defined by $\nabla V$. If we define the Hessian of a general configuration by
\be K_{i,\alpha;j,\beta} = \frac{\partial^2V}{\partial \vec{R_{i,\alpha}}\partial\vec{R_{j,\beta}}} \label{Eq:K} \ee
where $i,j=1,...,N$ and $\alpha, \beta \in \left\{x,y\right\}$, then the configuration will be stable if all eigenvalues of $K$ are nonnegative, while if there are negative ones (which correspond to imaginary frequencies of the normal modes), the configuration is unstable.

We emphasize that in this discussion, we are considering on equal footing all possible (stable and unstable) solutions, and not only the actual configuration that an ion crystal might take in an experiment. We are interested in exploring the structure of the space of crystal configurations, and for this we consider what solutions exist and how they are related for different values of the model parameter $\gamma_y$. Thus, we consider the solution space as a function of $\gamma_y$, but here we do not refer to the dynamics involved in an actual experiment which realizes a variation of the trapping parameters. The crystal configurations and also the eigenvalues of $K$ vary continuously with the model parameter $\gamma_y$. In general, when varying one model parameter, only one eigenvalue of the Hessian reaches zero at certain isolated values of the parameter. This is the case for all configurations considered here. Under these conditions, and for $\gamma_y > 0$, when an eigenvalue of the Hessian reaches zero, a bifurcation of the solutions occurs, with solutions either changing stability, getting `created' or `disappearing'. These bifurcations can be classified and the qualitative properties of the solutions on both sides of the bifurcation can be understood from considerations of symmetry and conservation of topological numbers, as we detail below.

For any solution with no zero eigenvalue, the local index of a solution is defined by $i_0\equiv {\rm sign}\det K=\left(-1\right)^n$, where $n$ is the number of negative eigenvalues. The sum of the local indices of all solutions emanating from the bifurcation point, is conserved separately on both sides of the bifurcations we consider, a result of topological degree theory \cite{Amann}. At each bifurcation, the soft mode whose frequency is equal to zero, is the mode which generates the spatial configurations after crossing the bifurcation point. The spatial symmetries of this mode determine the breaking of symmetry by the bifurcating solutions, as we detail below.

Fig.~\ref{Fig:Bifurcations} depicts schematically parts of the solution space of ion configurations for 31 ions and a given range of $\gamma_y$. There are many possible solutions which are not indicated, however most of them are unstable.
For $\gamma_y \gtrsim 152$ the only stable configuration is the linear chain along the $x$ axis. This configuration remains a solution for any value of $\gamma_y>0$, and its spatial structure is determined only by the number of trapped ions. Its axial and radial normal modes are exactly decoupled and the eigenvectors of the Hessian are independent of $\gamma_y$. In fact, the axial eigenvalues are also independent of $\gamma_y$ while the radial eigenvalues depend linearly \cite{Phonon_phonon_interactions} on $\gamma_y$. Therefore, as $\gamma_y$ is decreased below the critical value for stability of the linear chain solution, the radial normal modes of the chain lose stability one after the other at isolated bifurcation points, and these bifurcations generate new solutions in the solution space, as can be seen in fig.~\ref{Fig:Bifurcations} and fig.~\ref{Fig:Bifurcations50}.

%In fig.~\ref{Fig:Bifurcations}, the crystal configurations in the solution space are represented by curves to emphasize that they depend continuously on the model parameter $\gamma_y$, and so do the eigenvalues and eigenvectors of the Hessian. The only points where a noncontinuous dependence can occur are bifurcation points, where a crystal mode crosses zero frequency.
As the lowest-frequency radial mode of the linear chain crystal crosses zero (bifurcation \textit{a} in fig.~\ref{Fig:Bifurcations}), the linear chain solution bifurcates in a `pitch-fork' bifurcation, well studied in the thermodynamic limit \cite{MorigiFishman2008structural}. The linear chain becomes an unstable solution with one negative eigenvalue, and two new stable configurations appear in the solution space, the zigzag and its radial mirror image (`zagzig'), which become the new degenerate global minimum energy configurations.

The potential energy $V$ of eq.~\eqref{Eq:VNondimful} is separately invariant under two symmetry operations, inversion about the $x$ axis and inversion about the $y$ axis, symmetries which we denote as $\mathbb{Z}_2^x$ and $\mathbb{Z}_2^y$. The linear configuration has both symmetries. At the zigzag bifurcation, when the number of ions is odd (bifurcation \textit{a} in fig.~\ref{Fig:Bifurcations}), the radial symmetry $\mathbb{Z}_2^y$ is broken, but not the axial symmetry, and this directly results from the structure of the soft mode, depicted in the fig.~\ref{Fig:Bifurcations} as well. Since the chain solution is stable with index $i_0=\left(-1\right)^0=1$, and becomes unstable with $i_0=-1$ at the bifurcation point where a single eigenvalue crosses zero, the symmetry `protects' the spontaneously broken symmetric solutions which must have index 1 (since the sum of the local indices is conserved), and therefore are stable in all modes.

Following the unstable axial chain configuration to lower values of $\gamma_y$, the odd kink solution is created as the next eigenvalue of the axial chain crosses zero (bifurcation \textit{b} in fig.~\ref{Fig:Bifurcations}). Applied to the kink configuration, the axial and radial inversions $\mathbb{Z}_2^x$ and $\mathbb{Z}_2^y$ coincide and are individually broken, but not their combination (the kink is invariant only under total inversion). This is a result of the structure of the mode generating this bifurcation, which shows the same symmetry (for an odd number of ions). Since the chain solution is unstable with index -1 on the left of the bifurcation and a further mode turns negative, necessarily all bifurcating solutions are unstable. At bifurcation \textit{c} which occurs at lower $\gamma_y$, on each branch of the unstable odd kink solution, the remaining total inversion symmetry is broken, and this time the kink solution becomes stable, with the two bifurcating (necessarily unstable) solutions now breaking the symmetry.

Therefore we find that only for parameter values $\gamma_y \lesssim 106$, following bifurcation \textit{c}, can kinks exist as metastable excitations in the crystal. The centered kink solutions arise from a bifurcation of the axial chain solution, when the second-lowest mode crosses zero, and another bifurcation is required to endow the kink with stability. The unstable solutions of bifurcation \textit{c}, which are created in parallel with the kink solution gaining stability, are also of fundamental importance for understanding kink dynamics in the crystal. These unstable solutions are in fact `saddle-point' solutions, which separate a centred kink from a displacement by one lattice site \cite{PhysRevB.26.2886}. 

The kink solution which is stable at the centre of the crystal, cannot be found at any other site, niether stable or unstable - there is no kink configuration except at the centre for $\gamma_y \gtrsim 96.5$. The zigzag is not wide enough to support any other kink in the chain. Only at $\gamma_y \approx 96.5$ we find a bifurcation \textit{e} which creates a solution of an odd kink displaced one lattice site from the centre. Configurations of kinks displaced further away from the centre will be created at lower values of $\gamma_y$ via similar bifurcations.

Bifurcation \textit{e} is different from the pitch-fork bifurcations described previously, due to symmetry considerations. The solutions which `appear' for $\gamma_y \lesssim 96.5$ and do not exist for a larger $\gamma_y$, have none of the inversion symmetries of the potential and do not emanate from any solutions having such a symmetry. The local index to the left of the bifurcation is 0 (since no solution exists), and therefore two solutions are created and split apart in solution space, one is stable and the other unstable \footnote{Solutions with similar spatial structure, which are related by the two inversion symmetries, exist and are not shown explicitly in the figure}. Here as well, together with the (stable) displaced kink configuration, an unstable saddle-point configuration is created, which sets the barrier for the kink translation by one lattice site.

Following the saddle solutions of the centred odd kink configuration, bifurcation \textit{d} is reached at $\gamma_y\approx 45.6$. At this value of the parameter the crystal is wide, and the configuration and mode generating this bifurcation can be seen to have no inversion symmetries. Therefore, as in bifurcation \textit{e}, this is a bifurcation where the saddle-point meets a stable solution, and both disappear to the right of the bifurcation. The stable solution is a displaced `almost-extended' kink.
For $\gamma_y\lesssim 45.6$, the extended kink can only be found at the centre of the chain, and here we see that this occurs as the displaced extended kinks disappear by colliding with their saddle point solutions. We will elaborate on this phenomenon in the following section, where we look into the properties of the Peierls-Nabbaro potential in the setting of an inhomogenous crystal.
 
\begin{figure}[ht]
\center {\includegraphics[width=5.6in]{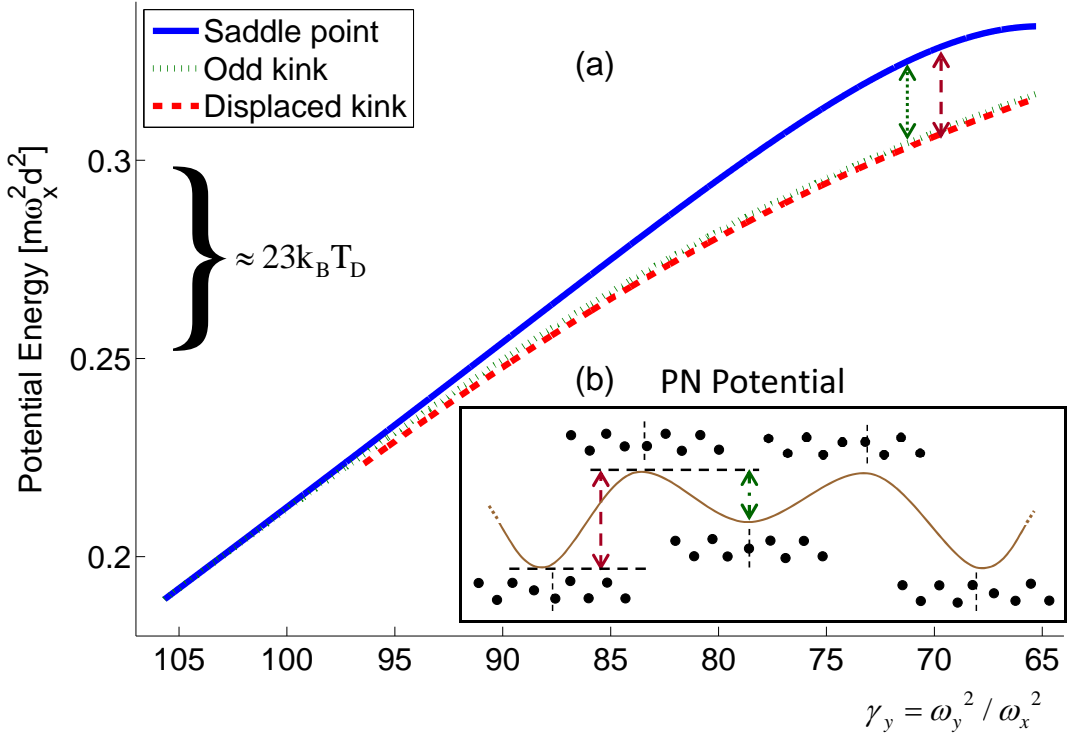}
\caption
{The potential energy of odd kink configurations in a planar crystal of 31 ions at different lattice sites, as in fig.~\ref{Fig:Bifurcations}. (a) The potential energy (in nondimensional units, see text) is plotted vs.~the model parameter, $\gamma_{y}\equiv \omega_{y}^2/\omega_{x}^2$. The energy of the zigzag (global minimum configuration) has been subtracted. The dotted green line can be used to define the rest mass of the odd kink at the centre of the crystal, in the entire parameter region where it is stable ($106\gtrsim\gamma_y\gtrsim 65.2$). The solid blue line is the energy of the saddle point configuration which separates the centered kink from the neighboring lattice sites (see bifurcation \textit{c} in fig.~\ref{Fig:Bifurcations}). The dashed red line gives the energy of the kink displaced by one lattice site. This configuration exists for $\gamma_{y}\lesssim 96$, see bifurcation \textit{e} in fig.~\ref{Fig:Bifurcations}. A measure of the potential energy in terms of the Doppler cooling limit temperature is given, for the trapping parameters of the experiment described in the text. (b) A schematic depiction of the three centre minima of the Peierls-Nabarro potential, which is the effective potential for (the motion of) a kink along the crystal. The minimum at the centre of the crystal is slightly higher, and the barrier for propagation to the next site on each side is obtainable from the saddle point configuration.}
\label{Fig:BifPotentials}}
\end{figure}

\section{Formation of the Peierls-Nabarro Potential\label{PNFormation}}

\begin{figure}[ht]
\center {\includegraphics[width=6in]{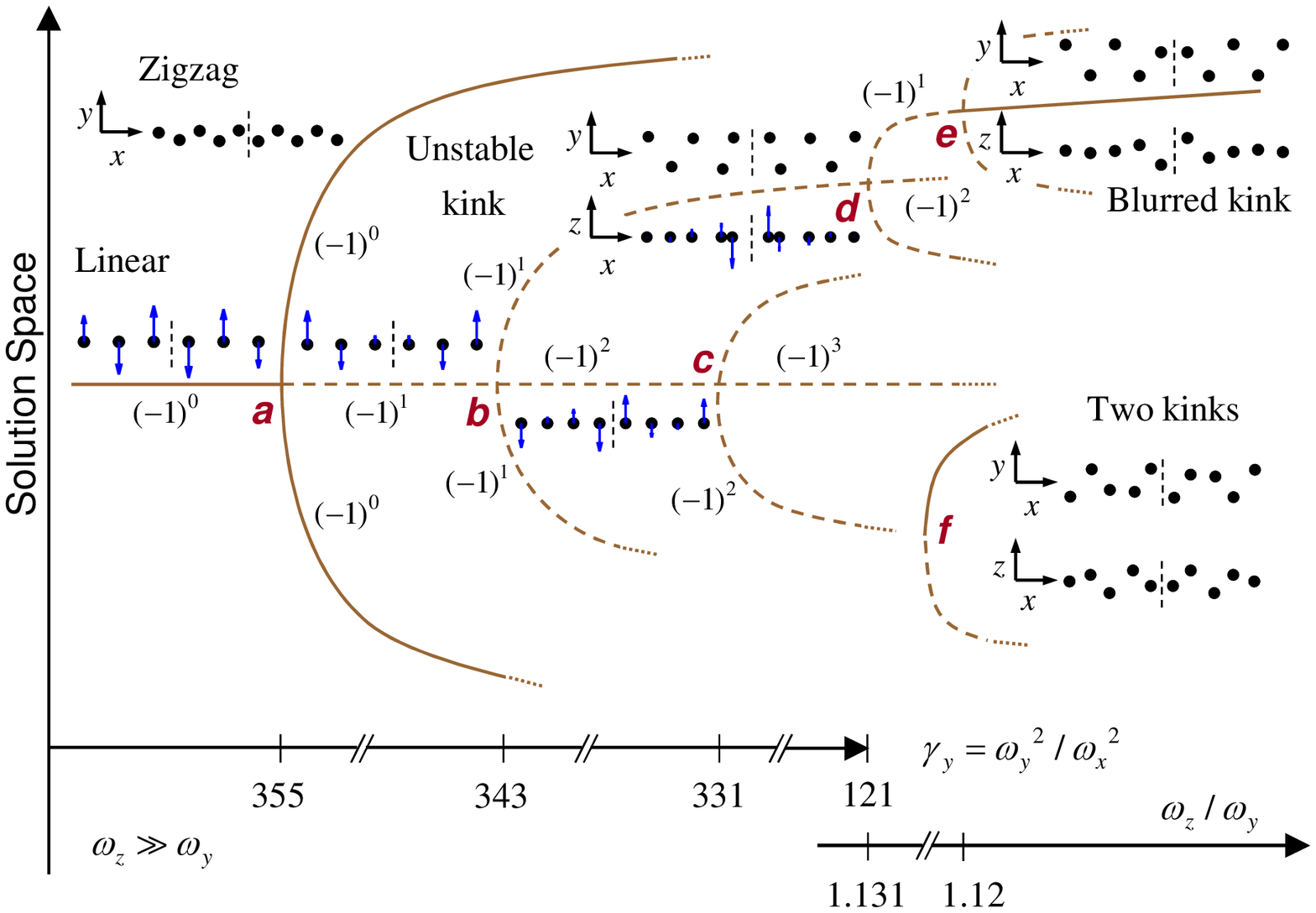}
\caption
{Bifurcations and stability of one- and two-kink configurations in a crystal of 50 ions. For three-dimensional crystals, there are two independent parameters which we take to be $\gamma_y\equiv \omega_y^2/\omega_x^2$ and $\omega_z/\omega_y$. On the left of the figure, we depict bifurcations as $\gamma_y$ is lowered, while $\omega_z\gg \omega_y$. On the right, bifurcations are shown for $\gamma_y=121$ held fix and $\omega_z/\omega_y$ lowered. For the notation in the figure see fig.~\ref{Fig:Bifurcations}. With the crystals which contain kinks, both the crystal plane ($xy$) and a side-view ($xz$) are shown, since the centre ions of the kink may extend out of the plane defined by zigzag crystal. See text for details of the bifurcations \textit{a}-\textit{f}.}
\label{Fig:Bifurcations50}}
\end{figure}

The saddle solution which bifurcates together with every stable kink configuration (see fig.~\ref{Fig:Bifurcations}) corresponds in fact to the Peierls-Nabarro (PN) potential barrier \cite{BraunKivsharBook} which separates the kink from neighboring lattice sites. Thus we see how the PN potential landscape forms in this model of an inhomogeneous discrete nonlinear system with a global trapping potential, through the same bifurcations in which the kink solutions are created. Moreover, examining the saddle solutions allows obtaining accurate quantitative properties of the PN potential barriers and local minima at each lattice site, as will be shown in the following.

Fig.~\ref{Fig:BifPotentials} shows the potential energy of 3 static kink configurations of 31 ions in a planar crystal as in fig.~\ref{Fig:Bifurcations}. The figure shows the dependence of the energy on $\gamma_y$, in the regime where the configurations are stable. The energy is measured in nondimensional units, corresponding to $V$ of eq.~\eqref{Eq:VNondimful}, and the energy of the global minimum zigzag configuration has been subtracted at each point. From the data in the figure, the landscape of the PN potential can be accurately defined and extracted (for the depicted centred and single-site displaced odd kink), as presented schematically in inset (b).  A measure of the energy in units of the Doppler-cooling limit temperature for the trapping parameters of the experiment is also given. 

For the odd kinks, which are defects contained within the zigzag boundaries, we see that in a large range of $\gamma_y$, the kink can be translated in the crystal, and successive bifurcations allow for stable kinks within local PN potential minima at lattice sites along the crystal. However, in some parameter regimes there exist stable solutions of kinks of different form, e.g.~the extended planar kinks, that can reside only at the centre of the crystal. This `trapping' of kinks at the crystal centre is the result of the PN potential being modified from a (modulated) periodic potential to a global trapping potential \cite{kink_trapping}. The PN potential is studied using a constrained energy minimization approach in \cite{TanjaPN}.

The centred extended kink is stable in the regime $63\gtrsim\gamma_y\gtrsim 25$. For $63\gtrsim\gamma_y\gtrsim 45$ it appears as depicted near bifurcation \textit{d} of fig.~\ref{Fig:Bifurcations} (experimentally imaged in fig.~1(f) of \cite{HaljanKinks}), while for lower $\gamma_y$ values it becomes more and more extended as the crystal becomes wider and wider (as in \cite{kink_trapping,TanjaKZ,HaljanKinks}).
The centred odd kink solution (of bifurcation \textit{c}) becomes unstable through a bifurcation which occurs at $\gamma_y\approx 65.2$. In this bifurcation the centre ion of the chain leaves the origin, breaking the last remaining composite symmetry $\mathbb{Z}_2^x \mathbb{Z}_2^y$, and the resulting kink configurations (stable in a small parameter regime $64 \lesssim\gamma_y\lesssim 65.2$) are similar to the (displaced) kink imaged in fig.~13 of \cite{Schneider2012}. In fact, although not directly connected by a bifurcation, these solutions are very close in shape to the `almost-extended' kink which is found for nearby parameters, as discussed above.

The transformations of different kink types as a function of the trapping parameters is an important property of kinks in Coulomb crystals. It has been used in \cite{TanjaKZ} to evolve odd kinks into extended kinks for stabilization, and investigated in detail also in \cite{HaljanKinks,TanjaPN}. It has been suggested in \cite{soliton_gate} to facilitate mobility of kinks in the lattice, since the mobility of kinks is enhanced near bifurcation points, with the translational mode becoming soft or the configuration becoming unstable, or even the PN potential landscape changing qualitatively.

In fig.~\ref{Fig:Bifurcations50} some configurations in the full solution space of 50 ions are considered, now without the restriction to planar crystals. Analyzing bifurcations which depend on two parameters is significantly more complicated than with one parameter, as simple bifurcations (with one mode crossing zero) occur generically along curves in parameter space. Bifurcations involving two modes simultaneously are also typical, occuring at points in two-dimensional paramete space. We restrict ourselves to just a few bifurcations leading to the formation of kinks, and vary only one parameter at a time.

At bifurcation \textit{a} the linear chain becomes unstable and the zigzag configurations are formed in the $xy$ plane, for $\omega_z>\omega_y$. Here, for an even number of ions, the zigzags are invariant only under total planar inversion, i.e. under the composition of $\mathbb{Z}_2^x$ and $\mathbb{Z}_2^y$. At bifurcation \textit{b} a centred kink is created. Now only $\mathbb{Z}_2^y$ is broken. This kink is not the odd kink configuration discussed above, because the odd kink would require a breaking of $\mathbb{Z}_2^x$ symmetry. We follow this unstable configuration down to $\gamma_y\approx 121$, while keeping $\omega_z\gg \omega_y$, thus restricting the crystal to the $xy$ plane. Now, lowering $\omega_z$ while keeping $\omega_y$ constant, at $\omega_z/\omega_y\approx 1.131$ bifurcation \textit{d} is hit, where the soft mode is transverse to the crystal plane (along the $z$ direction), breaking $\mathbb{Z}_2^z$.

The mode which generates bifurcation \textit{d} is a localized mode, with the eigenvector having significant components only on the centre ions of the kink. The bifurcating solution is no longer 2D (for $\gamma_y\approx 121$ held fixed, and $\omega_z/\omega_y<1.131$), but rather the kink-core ions extend out the crystal plane. This configuration becomes stable following a nearby bifurcation at \textit{e}, giving rise to the `blurred' kink solution found in \cite{kink_trapping}. The blurred kink at the centre of the crystal, is the only kink solution stable at this regime of parameters, for reasons which will be discussed below. Bifurcations \textit{c} and \textit{f} which are related to the formation of two-kink configurations, will be discussed in Sec.~\ref{TwoKinks}.

\section{Dynamics of Kinks in the Modified Peierls-Nabarro Potential\label{Dynamics}}

\begin{figure}[ht]
\center {\includegraphics[width=6.3in]{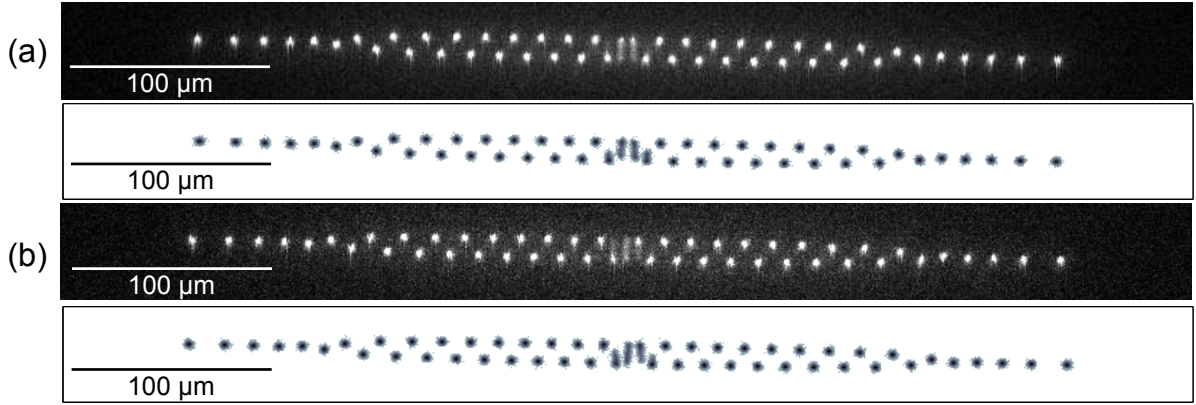}
\caption{Experimental and simulated integration images of a single kink at the crystal center, with (a) 50 ions, (b) 53 ions. In these simulations, the initial conditions correspond to the Doppler cooling limit temperature. For a total even number of ions in the trap, the kink appears symmetric about the centre. For a total odd number of ions in the trap, one of the kink ions appears more blurred than its immediate neighbour, which however is also distinctively more blurred than the ions outside of the kink core. The images reveal a $\sim -2^\circ$ tilt angle of the CCD camera with respect to the trap axis, that had been chosen for technical reasons, and treated exactly in the simulations (see App.~\ref{AppAnalysis}), and the crystals are nearly perpendicular to the line of sight, with the secular frequencies $\omega_x\approx 2\pi\cdot 56.7{\rm kHz}$, $\omega_{y}\approx 2\pi\cdot 623.3{\rm kHz}$ and $\omega_{z}\approx 1.047\omega_{y}$. The kinks are stable in the experiment for up to $\sim 10$s.}
\label{Fig:BlurredIntegrations}}
\end{figure}

In this section we extend the analysis of the `blurred' kinks which were first discovered and studied in \cite{kink_trapping}. The centre ions of these kinks appear blurred in the experimental images and this apparent blurring is reproduced in the numerical simulations using the trapping parameters of the experiment (see Fig.~\ref{Fig:BlurredIntegrations}). In order to derive the trapping parameters we reproduce the crystalline structures observed via a CCD camera, by using a molecular dynamics simulation and an optimization routine as detailed in App.~\ref{AppAnalysis}. The blurring results from large-amplitude and highly nonlinear oscillations of the centre kink ions, which correspond to an excitation of the kink's lowest-frequency localized mode. This large and anharmonic oscillation is a direct probe of the shape of the local PN potential which becomes globally trapping, with the kink-core ions extending out the crystal plane (see fig.~\ref{Fig:KinksModeProjection}).

As detailed in App.~\ref{AppAnalysis} (see also \cite{kink_trapping}), at the considered experimental parameters, $\omega_z/\omega_y\approx 1.047$. For this small radial indegeneracy, we find that the zigzag configurations starting with 53 ions are `quasi-3D' structures, with a few of the centre ions extending out of the zigzag plane in both directions, while the rest of the zigzag structure remains purely 2D. Since this transition starts with the ions at the centre, and the ions' change in position is small and nearly transverse to the crystal plane, the zigzags of 53-57 ions still seem to be planar zigzags when projected on the camera plane (which is almost parallel to the crystal plane). 
In addition, all kink configurations, starting with 46 ions, are also quasi-3D structures, with at least the kink-core ions extending out of the crystal plane, in opposite directions. This happens because a kink is a defect of increased ion density, which can be reduced by pushing the two kink ions, which are closest, oppositely out of the crystal plane, if the trap confinement in this direction is not strong enough.

\subsection{Linearized Normal Modes of Oscillations\label{NormalModes}}

\begin{figure}[ht]
\center {\includegraphics[width=6.3in]{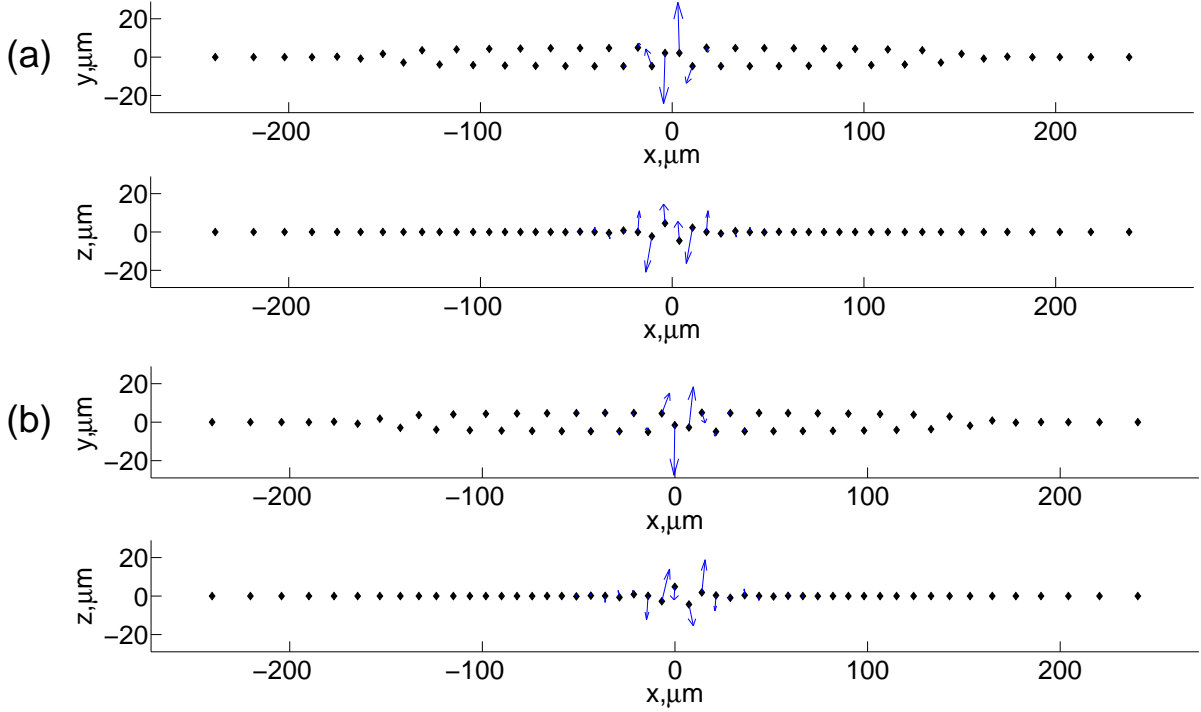}
\caption{The theoretically derived configuration of a kink in a crystal of (a) 50 ions, and (b) 51 ions, showing the crystal plane ($xy$ plane) and a side-view along the length of the crystal ($xz$ plane). The arrows on the center ions indicate the components of ion oscillations in the normal mode which is localized to the kink and has the lowest frequency (see text for details).}
\label{Fig:KinksModeProjection}}
\end{figure}

Fig.~\ref{Fig:KinksModeProjection} shows the configuration of a blurred kink in crystals of 50 ions and 51 ions, projected onto the crystal plane and transversally to it. It can be seen how the ions at the kink centre are pushed out of the crystal plane in opposite directions. In order to understand the dynamics of the kink configuration, we expand the potential of eq.~\eqref{Eq:VNondimful} to second order about the ion positions in a minimum-energy configuration, $\left\{\vec{R}_{i} ^{0} \right\}$, and obtain the normal modes $\Theta _{j}$ in a well known procedure (see e.g. \cite{KinkCoherence}), by setting 
\be\vec{R}_{i,\alpha} \left(t\right)=\vec{R}_{i,\alpha} ^{0} +\sum _{j}^{3N}D_{i,\alpha}^{j} \Theta _{j} \left(t\right),\label{Eq:R}\ee where $D_{i,\alpha}^{j} $ is the matrix of normal mode vectors, with rows indexed by the $N$ ion numbers $i$ and $3$ directions $\alpha$, and columns by the $3N$ normal mode numbers $j$.

\begin{figure}[ht]
\center {\includegraphics[width=6.3in]{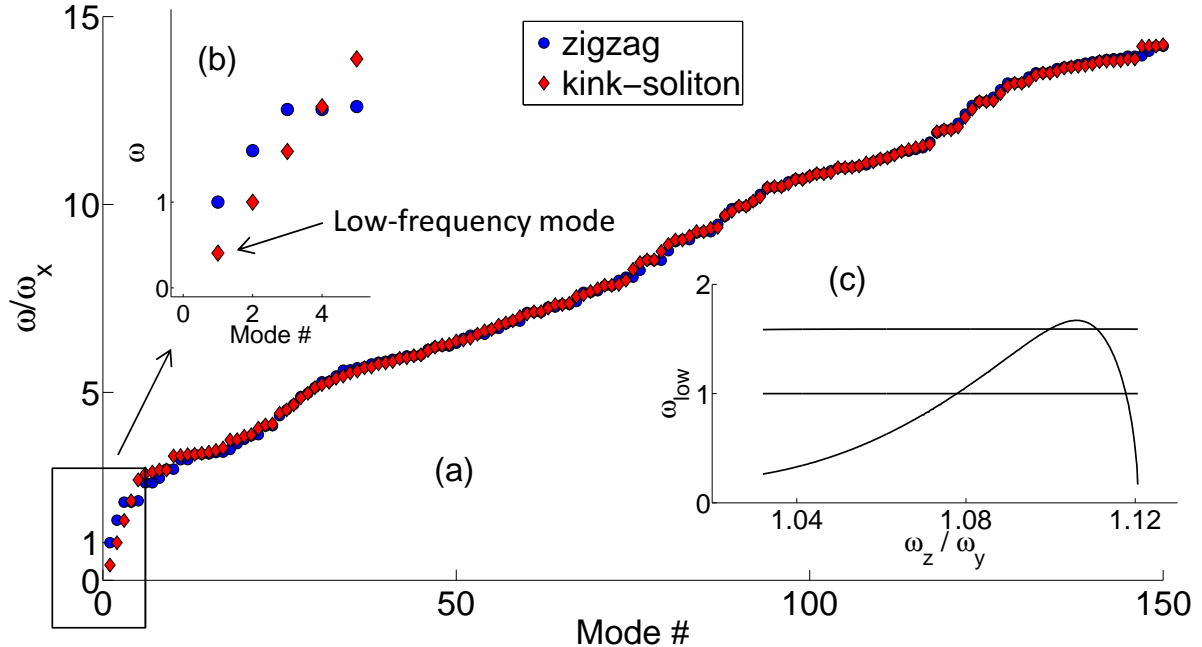}
\caption{(a) The frequency spectrum (in units of the axial c.o.m frequency, $\omega_x\equiv 1$) of normal modes of the zigzag (blue circles) and the kink (red diamonds), of the configuration of Fig.~\ref{Fig:KinksModeProjection}a, with 50 ions. (b) The low-frequency localized mode of the kink lies below the axial c.o.m frequency, which is the lowest mode in the zigzag spectrum, as can be seen in the inset. (c) Dependence of the kink's low-mode frequency ($\omega_{\rm low}$) on the radial indegeneracy. Near the left and right limits of the inset, the low-frequency mode approaches zero, which signifies loss of stability of this kink configuration. }
\label{Fig:KinkDispersion}}
\end{figure}

The frequency spectrum of normal modes of the kink configuration of Fig.~\ref{Fig:KinksModeProjection}a, with 50 ions, is shown in Fig.~\ref{Fig:KinkDispersion}a, where also the spectrum of the 50 ion zigzag is plotted for comparison. The spectra of the two configurations overlap for the most part (there are differences in the normal modes). However, the notable localized low-frequency kink mode, which has in this case frequency $\omega_{\rm low}\approx 0.4\omega_x$, lies well below the lowest frequency of the zigzag configuration. The components of ion oscillations in this low-frequency mode vector are depicted by the arrows in Fig.~\ref{Fig:KinksModeProjection}a. The lowered frequency of this mode is a result of the near-degeneracy of the two radial trapping frequencies, and the motion of ions corresponds to oscillations in directions for which the restoring forces are minimal - i.e. which are `empty' of ions, pointing away from the locally high density of the defect.

For the kink with 51 ions in fig.~\ref{Fig:KinksModeProjection}b, it can be seen how due to the small radial indegeneracy, the kink-core ions are pushed out of the crystal plane in slightly asymmetric directions, with the localized low-frequency normal mode components expressing the assymetry in the configuration. This explains the different blurring pattern of the kinks with an even and odd number of ions as observed in the experiment (Fig.~\ref{Fig:BlurredIntegrations}).

In Fig.~\ref{Fig:KinkDispersion}c the dependence of the localized low-mode's frequency is plotted vs. the radial indegeneracy, $\omega_z/\omega_y$. Structural transitions occur at $\omega_{\rm low}=0$, where this kink configuration loses stability. For intermediate values of the indegeneracy, $\omega_{\rm low}$ increases and can surpass the axial c.o.m mode's frequency (constant at $\omega_x\equiv 1$), and even the second axial mode.

We note that for the low-frequency mode of the kink, there may be some deviations in frequency, introduced by the affect of micromotion \cite{rfions,rfmodes,zigzagexperiment}, as compared with the pseudopotential analysis presented here. We have verified that these deviations do not alter the general results of the analysis, except for some fine tuning of the parameters, of the order of a few percent \cite{zigzagexperiment}, depending on the exact configuration.

\begin{figure}[ht]
\center {\includegraphics[width=5.0in]{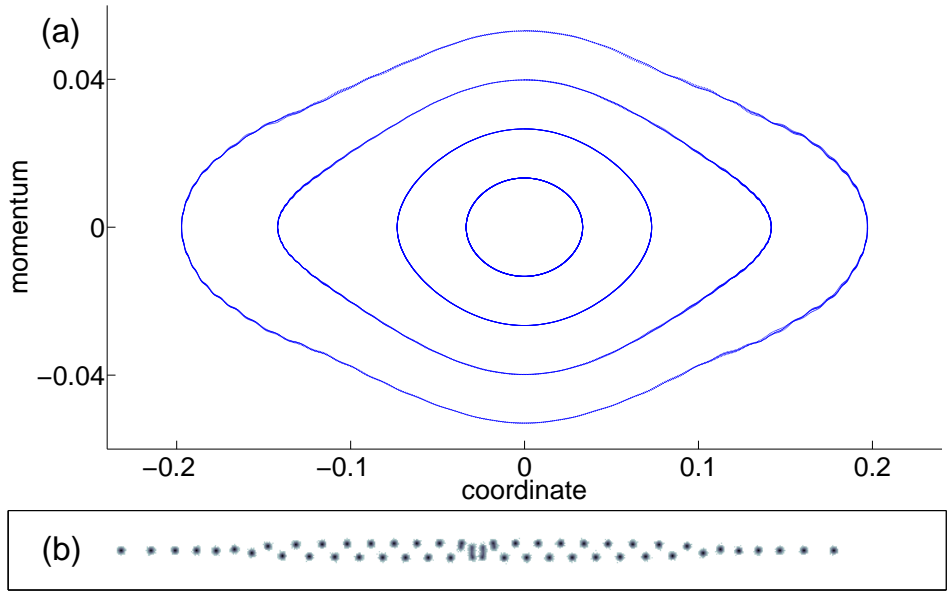}
\caption{(a) Orbits in phase-space (nondimensional units, see text) of the oscillator corresponding to the low-frequency kink mode, obtained by plotting four different runs of the dynamical simluation, with increasingly larger initial excitation, only of the low frequency mode. The outer most orbit corresponds to an initial energy of $\sim 2/3$ of the Doppler cooling limit temperature, T$_{\rm D}$. The low-energy orbits are nearly elliptical, as expected for harmonic motion, while the motion at higher energies is clearly anharmonic. (b) An integration image corresponding to the outermost orbit, demonstrating that the typical blurring observed in the experiment can be explained as an excitation of the low-frequency mode.}
\label{Fig:PhaseSpace}}
\end{figure}

\subsection{Nonlinear Oscillations\label{Nonlinear}}

The strong Coulomb nonlinearity which creates the kink and enables the high ion density at the kink center, means that the description of ion oscillations in terms of linearized decoupled modes will break at a certain amplitude of motion, when the nonlinear terms become significant. It was shown in \cite{KinkCoherence} that different localized modes are coupled much stronger than is typical in ion crystals. The blurring of the kink ions, observed in the experiment and in numerical simulations which were run at the Doppler cooling limit temperature (T$_{\rm D}$), corresponds to a sufficiently large amplitude motion of the kink ions, to turn the description in terms of a linear oscillator (of the low frequency normal mode) inadequate.

An exact numerical simulation of the motion of the ions (in the potential of eq.~\eqref{Eq:VNondimful}) was run, with initial conditions corresponding to an excitation only of the localized low-frequency mode of the kink, within the pseudopotential approximation. Fig.~\ref{Fig:PhaseSpace} shows phase-space orbits of the oscillator corresponding to the low-frequency kink mode (obtained by inverting eq.~\eqref{Eq:R}), for 4 different runs with increasingly larger motional energy. Still, the outer most orbit corresponds to an initial excitation energy (only for the low-frequency mode) of $\sim 2/3$ of T$_{\rm D}$. The low-energy orbits are nearly elliptical, corresponding to harmonic motion. However, the oscillator is clearly anharmonic for motion amplitudes which are relevant for the experiment, with a temperature in the range T$_{\rm D}$ to 5T$_{\rm D}$. Fig.~\ref{Fig:PhaseSpace}b shows an integration image corresponding to the outermost orbit, demonstrating that the typical blurring observed in the experiment can be obtained solely as excitation of the low-frequency mode, in a dynamical simulation without micromotion or laser-related effects. It should be noted that this does not mean that at T$_{\rm D}$, the anharmonic normal mode remains an uncoupled oscillator - it is coupled to other degrees of freedom of the chain. However, when $\omega_{\rm low}$ is tuned to relatively high value (as in the middle region in fig.~\ref{Fig:KinkDispersion}c), which is an indication of the local potential turning stiffer, the large amplitude anharmonic motion is suppressed and the blurring of the kink ions disappears in the simulations.

\section{Kinks with a Mass Defect and Two-Kink Configurations\label{TwoKinks}}

\begin{figure}[ht]
\center {\includegraphics[width=6.4in]{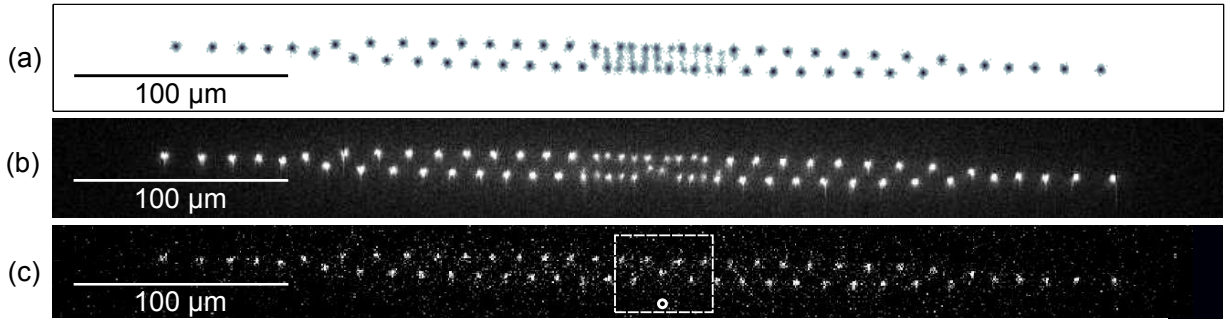}
\caption{Numerically simulated and experimentally observed integration images of solitons (kinks) oscillating within the PN potential provided by the Coulomb crystal. (a) Simulated integration image of a kink which is `released' from rest, six lattice sites to the side of the trap centre, with 56 ions. The large area of blurring reveals that the kink oscillates along the crystal. In these simulations, no cooling was assumed, and the crystal is initialized at `zero' temperature. The image was integrated for $\sim 1$ms. In the experiment, such rapid kink oscillations would quickly damp, both by the cooling laser and by the radiation of phonons from the kink. (b) Experimental observation of a kink, additionally incorporating a mass defect, both hopping within the crystal for about 6 seconds. (c) A single image (200\,ms) from the integrated image in (b), showing 57 bright ions. We deduce the existence of a dark, heavy molecular ion from the numerical fit as described in the text (see also fig.~\ref{Fig:Experiment3D}).}
\label{Fig:KinksOscillating}}
\end{figure}

In the numerical simulations, blurred kinks which are created displaced from the centre of the crystal, immediately start to oscillate about its centre. Fig.~\ref{Fig:KinksOscillating}a shows a simulated integrated fluorescence image of a kink `released' from a few lattice sites to the side of the centre. We note that in this simulation the system is initialized at `zero' temperature, and the Doppler cooling is neglected. The blurring across the lattice reveals the oscillation of the kink along the chain. In the experiment, events like this require a careful choice of the parameters, since the motion gets easily damped, both by the cooling laser and by the radiation of phonons from the kink \cite{BraunKivsharBook} as it moves in the PN potential.

%Massive kinks 
Experimental evidence of a structural defect hopping back and forth along the crystal on a timescale of $\sim$ 100\,ms/step is shown in fig.~\ref{Fig:KinksOscillating}b, where the image consists of an integration accumulating light scattered during $\sim$~6 seconds. 
A single CCD-image (exposure 200\,ms) is shown in fig.~\ref{Fig:KinksOscillating}c, revealing the contribution of a different, dark ion species to the kink. From exact numerical simulations including the time-dependent potential we deduce a dark ion mass of at least 40 atomic mass units. This could be explained by collisions with background gas ($\sim$\, 0.1/sec), such as ${\rm Mg}^+ + {\rm H}_2{\rm O} \rightarrow {\rm Mg}^+\left( {\rm H}_2{\rm O} \right)$, due to a vacuum around $10^{-9}$\,mbar \footnote{Different dark ion species of mass/charge ratio similar to the ${\rm Mg}^+$ ions are also observed, however these get directly embedded into the Coulomb crystal, causing lattice sites not featuring flourescence light}. We observe melting events of the crystal, most likely induced by a collision with the residual gas, followed by the occurrence of a dark ion. Since we find that the dark ion has a high probability to turn into a bright ion after a subsequent collision, we conjecture it to be a molecular ion incorporating a Mg$^+$ ion. The lack of closed electronic transitions in such ions explains that they are not directly observable via flourescence light. However, the dark ion gets sympathetically cooled by its coupling to the cold bath of already laser cooled Mg$^+$ ions via mutual Coulomb interaction. The different mass/charge ratio causes it also to experience a reduced confinement by the mass/charge dependent pseudopotential as compared with the Mg$^+$ ions. As a consequence, the Coulomb repulsion additionally displaces the molecular ion out of the zigzag crystal to roughly twice the radial extent of the other ions, so in our case to $\sim 10\mu$m away from the trap axis. Furthermore, this displacement occurs also out of the crystal plane, therefore deforming the structure of the Coulomb crystal, either combined with a kink or as a configuration which corresponds to a groundstate distorted zigzag.

%We identify ${\rm MgH}^+$ or $^{25}{\rm Mg}^+$  dark ions as they take almost exactly the position of ordinary ${\rm Mg}^+$ ions. However, we also see crystals with a dark large-mass defect, where due to the weaker radial trapping for the heavier ion, it is located far outside of the zigzag,

\begin{figure}[ht]
\center {\includegraphics[width=6.4in]{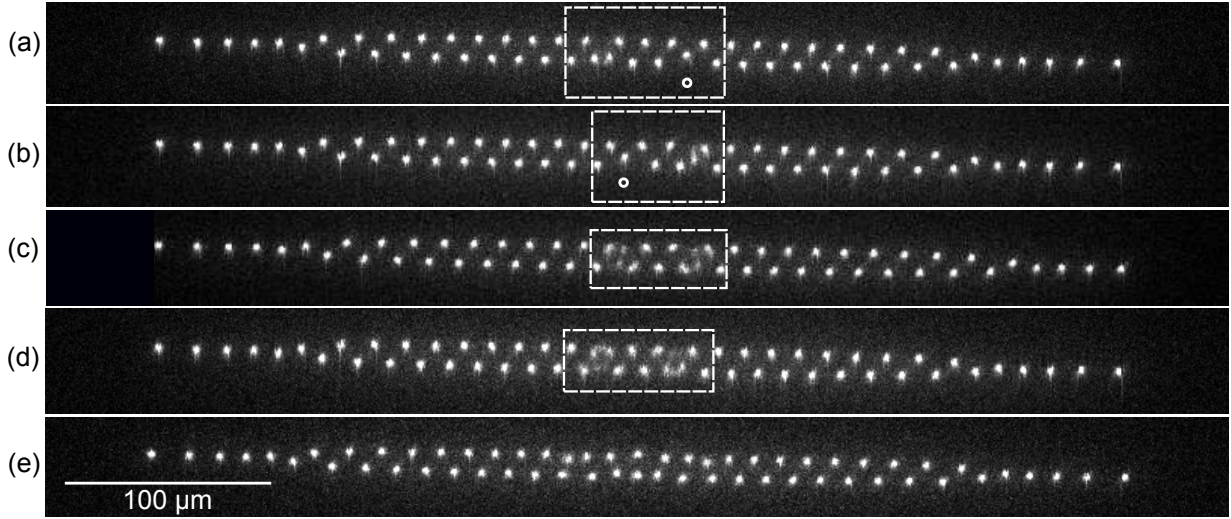}
\caption
{Experimental CCD images of Coulomb crystals with different numbers of ions incorporating different species of kinks, even more then one at a time. (a)-(d) Crystals containing 58 ions with two kinks, highlighted by dashed white boxes. The kinks can no longer be centred in the crystal due to their repulsive interaction at sufficiently low internal temperature. Images (a) and (b) show a kink incorporating a mass defect (the dark ion is depicted with a white circle), in combination with an even (a) or an odd (b) blurred kink. The crystals shown in (c) and (d) contain two kinks of the same type. These configurations are stable for several seconds. In (e) the central region depicts how larger number of ions (here 60 Mg$^+$) form a complex helical structure, projecting these structures to the CCD plane these wavelike ion chains are obtained.}
\label{Fig:Experiment3D}}
\end{figure}

The hopping of the massive kink in the crystal of fig.~\ref{Fig:KinksOscillating}b occurs on a long timescale, and characterizing the exact dynamics requires further studies. The controlled vibrational excitation of the massive kink exploiting its again altered and gapped spectrum in the PN potential and a fast imaging setup will allow us to study the kink with a mass defect and the related dynamics \cite{HaljanHopping,TanjaPN}. In addition, the kink with a mass defect occurs also in combination with a second kink in the crystal (fig.~\ref{Fig:Experiment3D}a-b), which leads us to consider configurations of two kinks \cite{MainzKZ,HaljanKinks,TanjaPN} and their interaction.

%Two-kink configurations
Returning to the bifurcation diagram of 50 ions in fig.~\ref{Fig:Bifurcations50}, at bifurcation \textit{c} the third-lowest mode of the axial chain crosses zero and generates a solution which is not invariant under inversions about $x$ or $y$, but only under the combined operation. This solution is planar, incorporating two kinks, however it is unstable. The solution of two kinks side by side, which are stable at the experimental trapping parameters, is predicted after bifurcation \textit{f}, where the $z$-inversion symmetry is broken, and the two solutions are not connected at the shown `cut' through parameter space. An isolated simple bifurcation (with one mode crossing zero) connecting these two solutions is ruled out by considerations of index conservation (see Sec.~\ref{Bifurcations}) and the single $z$-inversion symmetry which must be broken.

Moreover, from calculations similar to those related to fig.~\ref{Fig:BifPotentials}, it is possible to find the energy gap compared to the groundstate of the two-kinks configuration. For 50 ions in the experimental parameters $\gamma_{y}=11^2$ and $\omega_{z}/\omega_{y}=1.047$, it is E$_{\rm two-kink}=0.27$, which should be contrasted with $2{\rm E}_{\rm kink}=2\cdot 0.1265=0.253$, where the energy of the single kink is taken at the centre. Kink configurations with a larger separation can also be stable, and the dependence of the interaction energy of the kinks on their distance can be computed. Also of importance is the energy barrier for collision of the two kinks (which results in their annihilation). Using the bifurcation method presented in this paper, the barrier can be exactly defined and is computed to be E$_{\rm barrier}=0.078$, which corresponds for the experiment to $\sim 35{\rm k}_{\rm B}$T.

Therefore we find numerically and experimentally that at a sufficiently low temperature, which renders improbable high amplitude oscillations leading to annihilation, two kinks can exist side by side, repelling each other and stabilized at a defined position by their interaction. Fig.~\ref{Fig:Experiment3D}c-d shows two such experimentally realized configurations.

Finally, the interaction of a kink with a mass defect and different blurred kinks (see fig.~\ref{Fig:Experiment3D}a-b) also results in a change of the local potential landscape. In the case shown here, the presence of the massive kink shapes the local PN potential. As a directly detectable result, the kink no longer appears blurred, an indirect evidence for the interaction of the two kinks causing an additional modification of the local PN potential. Along the lines of Sec.~\ref{Bifurcations}-\ref{PNFormation}, a study of the PN potential bifurcations for such configurations can be performed, for example by taking the mass of the defect ion to vary continuously as another parameter. In addition the spectrum of normal modes of two interacting kinks can be studied as in Sec.~\ref{Dynamics}. The localized modes of the two kinks will interact and allow for exchange of energy between these two kinks as well as with the rest of the mesoscopic ensemble of ions and collective modes of the crystal. This might become exploitable to study open quantum system dynamics \cite{soliton_gate}.

\section{Concluding Comments\label{Conclusion}}

To conclude, the concepts introduced in this paper, revealing the bifurcations of configurations in inhomogenous trapped ion crystals, have enabled us to give an unambigous definition the PN potential landscape for one or more kinks. By following the bifurcations of relevant solutions in dependence on the trapping parameters, the existence and stability of different kink solutions can be predicted. We experimentally observe and numerically confirm how for different trapping parameters and related configurations, the kinks can be either highly mobile, pinned locally at lattice sites along the crystal, or even trapped globally by a modified PN potential. Variation of the trapping parameters in a dynamical way allow for different kinks to be transformed into each other, or set to motion along the crystal. Controlled manipulation of kinks in Coulomb crystals will enable further studies into the dynamics of discrete solitons with unmatched experimental accessibility.

Studying the localized modes of kinks is a further valueable tool for understanding kink dynamics, both linear and nonlinear. Depending on the trapping parameters, kinks may have localized modes which for cold crystals, can be well described in terms of linearized small oscillations, while for some configurations, the nonlinearity may play a crucial role even for laser-cooled crystals. The isolation of such collective degrees of freedom, or, on the other hand, the coupling of these nonlinear modes to their environment, is of fundamental interest. As studied in detail in \cite{KinkCoherence,soliton_gate}, cooling one or a few modes close to the quantum groundstate may be feasible (in particular with the odd kinks, which manifest a gap separated high-frequency internal mode). Therefore the ion trap system may allow reaching the quantum regime for motional modes localized to unique topological solutions, together with the ability to accurately measure the dynamics of these modes. This could allow for a variety of novel experimental studies - starting from open quantum system dynamics, to using solitons for conveying quantum information. In addition, quasi-2D zigzag phases similar to those realized in the ion trap system, have recently also attracted attention in the context of dipolar gases \cite{PhysRevA.78.063622,PhysRevB.85.125121} and 1D quantum wires \cite{PhysRevLett.98.126404,0953-8984-21-2-023203,PhysRevLett.110.246802} and the study of discrete solitons in such systems may be an intriguing further step. 

The configurations presented here of two interacting kinks, either of the same type or combined with a mass defect, are a first step towards studying the interaction of solitons. We have shown that two-kink configurations realizable in the ion trap are stable for times which are orders of magnitude longer than the natural timescale of the dynamics in the system, and the barrier for annihilation of two solitons can be computed accurately, as has been demonstrated for a specific case. We also demonstrated that the interaction between kinks may cause a modification of the PN potential in their vicinity, leading to observable effects in their motion. Various promising possibilities exist for pursuing studies of kink interactions, first classically and eventually perhaps quantum mechanically, in this rich and complex nonlinear system.

\begin{acknowledgments}
TS was funded by the Deutsche Forschungsgemeinschaft (SCHA973). MM, TS and BR acknowledge the support of the European Commission (STREP PICC). BR acknowledges the support of the Israel Science Foundation and the German-Israeli Foundation. HL, TS and BR thank A. Retzker, T. Mehlst{\"a}ubler and  M. B. Plenio for fruitful discussions, held in particular during the ``Ulm Workshop on Theoretical and Experimental Aspects of Nonlinear Physics in Ion Traps''. HL thanks R. Geffen.
\end{acknowledgments}

%%%%%%%%%%%%%%%%%%%%%%%%%%%%%%%%%%%%%%%%%%%%%%%%%%%%%%%%%%%%%%%%%%%%%%%%%%%%%%%%%%%%%%%%%%%%%%%
\appendix

\section{Analysis of the Experimental Results}\label{AppAnalysis}

In this appendix we describe the exact form of the time-dependent trapping potential modeling the trap used in the experiments, and formulate how the orientation of the crystal plane can be calculated, taking into account the time-dependence of the solution. The fitting routines which allow the accurate analysis of crystal structures and trap parameters are described. The calculations presented here were also used to derive the structure of crystals in \cite{kink_trapping}.

\subsection{A Tilted Trap Potential\label{TrapPotential}}

We start with the trapping potential of a linear Paul trap \cite{WinelandReview}, taking $x$ as the axial direction, and (in this section) measuring time and distances in units of $2/{\Omega}$ (where ${\Omega}$ is the rf frequency) and $d=\left(e^{2} /4m{\Omega}^{2} \right)^{1/3}$, respectively (with $m$ and $e$ the ion's mass and charge). The nondimensional trapping potential for a single ion is
\be {V}_{\rm 4-rod} =\frac{1}{2} \left(a_{x} x ^{2} +a_{y} y^{2} + a_{z} z^{2} -2q\cos2t\left(y^2 -z^{2} \right) \right) , \label{Eq:Vtrap1}\ee
with $q$ and $a_{\alpha }, \alpha \in \left\{x,y,z\right\}$, being the nondimensional Mathieu parameters of the respective coordinates \cite{WinelandReview}.
Since the Laplace equation requires that $ \sum _{\alpha }a_{\alpha }  = 0 $, we take $a_z = -a_x - a_y$. 

To the potential of eq.~\eqref{Eq:Vtrap1} we add further harmonic terms pertaining to a DC trapping voltage in the radial plane, however whose principal axes are different, i.e. 
\be V_{\rm rot} = \frac{1}{2} \left(a'z'^2 - a'y'^2\right) \label{Eq:Vrot}\ee
where the $y'z'$ plane is rotated with respect to the $y$ and $z$ coordinates by some angle $\theta$, and $V_{\rm rot}$ obeys the Laplace equation by construction. Writing $y',z'$ in terms of the original coordinates, we get the trapping potential in the form
\be {V}_{\rm trap} = \frac{1}{2} \left(a_{x} x ^{2} +a_y y ^{2} +a_{z} z ^{2} +a_{yz} y z - 2q\cos2t\left(y^2 -z^{2} \right) \right)\label{Eq:Vtrap} \ee
where the parameters have changed by 
\[ a_y \to a_y -\left(\cos^2\theta - \sin^2\theta\right)a', a_z \to a_z +\left(\cos^2\theta - \sin^2\theta\right)a'\]
 and we have $a_{yz} \equiv 2a'\cos\theta \sin\theta $. We note that for $\theta=45^\circ$, the original parameters $a_y$ and $a_z$ are unchanged, and $a_{yz}\equiv a'$.

For a single cold ion trapped at the origin of the potential $V_{\rm trap}$ of eq.~\eqref{Eq:Vtrap}, the parameters of the potential will affect the frequencies and eigenvectors of its modes of oscillation about the origin. It is instructive to consider a few simple cases. The $x$ motion is of course decoupled and depends only on $a_x$. Setting first $a_{yz}=0$, the motion in the two radial directions (along the $y$ and $z$ axes) decouples and the modes will be aligned with those axes. For $a_y=a_z=-a_x/2$, the two radial frequencies will be exactly degenerate. This degeneracy can be lifted by varying $a_y$ (while $a_z$ is not an independent parameter, but determined by $a_z = -a_x - a_y$). Setting $a_{yz}\ne 0$ couples (linearly) the motion in the radial plane, causing the mode vectors to mix, such that their direction is no longer aligned with the $y$ and $z$ axes \cite{Electrostatics_of_surface_traps,SurfaceTrapLucas,Stability_of_surface_traps}. The frequencies of the radial modes will change as well. The magnitude of the radial modes' indegeneracy, and the rotation of the principal axes, depend also on the value of the other parameters ($q$, $a_y$ and $a_z$) in a complicated manner.

With crystals of multiple identical ions, the effects described above hold for the three center-of-mass (c.o.m) modes, which are decoupled from other modes of the crystal. The pseudopotential approximation is constructed by taking the trapping potential to be time-independent, with principal axes defined by the orthogonal directions of the c.o.m modes, and trap strength given by the respective frequencies . In many cases, the structure of the crystal which is predicted within the pseudopotential approximation, is very close to the crystal which is a solution of the full time-dependent potential \cite{zigzagexperiment}. However, it is known that there exist regions in parameter space, in which crystals may be found in `peculiar' configurations, which do not exist within the pseudopotential approximation \cite{Blumel1995,Brewer1995,Drewsen2000,rfions}. The modes of oscillation of the ions in the time-dependent potential are in many cases also very close to the modes calculated from a pseudopotential, but there may be corrections to the eigenvectors or eigenfrequencies, which can be calculated \cite{rfions,rfmodes,zigzagexperiment}. In the cases examined here, the time-dependence of the potential results in some tuning of the parameters, which was accurately accounted.

\subsection{Fitting the Trapping Parameters\label{TrapFitting}}

We fit the experimental trapping parameters using an optimization routine which works in the following way. The free parameters in $V_{\rm trap}$ of eq.~\eqref{Eq:Vtrap} are taken to be $a_x,a_y,a_{yz}$ and $q$. The theoretical crystal configuration for given trapping parameters is obtained by a molecular dynamics simulation of the equations of motion of the ions with some small damping, which is adiabatically turned off. The resulting solution is verified to be a stable crystal configuration. If the initial conditions for the simulation correspond to a chain, the zigzag configuration is easily obtained. Finding a kink configuration requires starting with a zigzag configuration in which the radial coordiantes of half the crystal ions were flipped.

\begin{figure}[ht] \center {\includegraphics[width=6.4in]{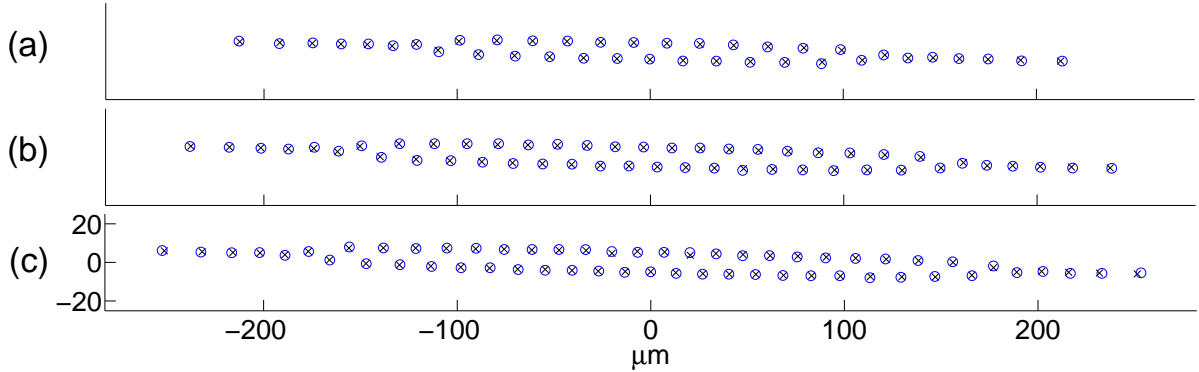} \caption{Comparison of numerically obtained configurations to the experimental data, with (a) 39 ions, (b) 50 ions, (d) 57 ions. Theoretical average ion locations, as projected on the camera plane (see text for details), are shown with crosses, and the experimentally imaged ions are donated with circles. The average deviation between the experimental data and theory is smaller than $0.5\mu m$ per ion, which is less than the pixel size ($0.8\mu m$) and well within $1\sigma$ for all ions.} \label{Fig:ZigzagProjections}} \end{figure}

The theoretical crystal at given parameters is then compared to the experimental image, by the following steps. The coordinates of all the bright ions are extracted from the experimental image. Then the ion coordinates are measured from the center-of-mass, which coincides with the center of the trap for crystals with no dark ions. For crystals with dark ions, the center is taken to be the average of several all-bright configurations. The orientation of the camera plane with respect to the trap radial axes is parameterized by two additional free parameters, the azimuth and elevation angles, which are known to be close to $0^\circ$ and $-45^\circ$. The theoretical crystal's 3D coordinates are projected onto the plane of the camera, and then the sum of distances between each ion in the experimental image and the theoretical projection is computed. A least-squares-fit is performed to minimize this sum in order to find the trapping parameters in combination with the point of view of the camera which best reproduce the experimental images.

Running the above analysis for a single crystal image does not determine the parameters uniquely. However, already for the crystal with 39 ions (Fig.~\ref{Fig:ZigzagProjections}a) it is immediately found that the crystal plane cannot coincide with the trap's $y=0$ or $z=0$ planes, which are at $\pm 45^\circ$ to the camera, since in this case, the width of the crystal is $\sqrt{2}$ wider than appearing in the image, and a matching configuration does not exist. Quantitatively however, the image could be generated by different combinations of trapping parameters, which correspond to a crystal plane which is either facing the camera (i.e. lying in the $y=-z$ plane), or rotated from it at up to $\sim 22^\circ$. A nonzero $a_{yz}$ is necessary for reproducing the observed crystals, in a plane which is viewed top-on from the camera. [A zigzag crystal which lies in $y=-z$ plane can be reproduced in the simulations by using nearly degenerate radial trapping freuqencies, which result in a `peculiar' planar zigzag in the $y=-z$ plane for 39 ions (the slight indegeneracy which is still present distinguishes this plane from the $y=z$ plane). However, most of the other crystals which were observed in the experiment could not be explained as `peculiar' crystals, since the observed configurations do not exist or are unstable when the radial frequencies are near-degenerate - they all require that the out-of-plane trapping frequency be larger then the in-plane one, by at least a few percent.]

In the experimental data, images of a zigzag configuration are found with up to 57 ions. For 58 and 59 ions, there is a variety of images of configurations with defects, and starting with 60 ions the configurations in the 2D projection seem `wavey' (Fig.~\ref{Fig:Experiment3D}e). These crystals can be reproduced in the simulations as various three-dimensional structures. Trying to fit unique trapping parameters for the different crystals with between 39 and 57 ions, it is found that the outer most ions of the longest crystals observed, are a bit too expanded sideways, an effect which can be explained by a small ($\sim 0.5\%$) decrease in the axial trapping on the sides, due to nonlinearity effects. Extensive searches in parameter space using dynamical simulations, show that all of the observed images can be accounted for quantitatively and the trapping parameters are derived. The average deviation between the experimental data and theory is smaller than $0.5\mu m$ per ion, which is less than the pixel size ($0.8\mu m$) and well within $1\sigma$ for all ions. The following parameters were obtained from the theoretical fitting : $a_x=0.000328$, $a_y=-0.0002$, $a_{yz}=0.0019$ and $q=0.286$. The rf frequency from the experiment is $\Omega=2\pi\cdot 6.22{\rm MHz}$, and the ions are $^{24}{\rm Mg}^+$. The camera plane is found to be at azimuth angle of $\sim -1.92^\circ$ and the elevation $\sim-44.5^\circ$. The crystals lie almost exactly in the $y=-z$ plane (in rotated coordinates $y'=0$), and the secular frequencies are $\omega_x\approx 2\pi\cdot 56.7{\rm kHz}$, $\omega_{y'}\approx 2\pi\cdot 623.3{\rm kHz}$ and $\omega_{z'}\approx 1.047\omega_{y'}$.

%\begin{acknowledgments}
%\end{acknowledgments}

\bibliography{kink_bifurcations}

\end{document}